\documentclass{emulateapj} 

\begin{document}
\shorttitle{The Cygnus x-1 jet} \title{Composition,
collimation, contamination: the jet of Cygnus X-1}

\shortauthors{Heinz} \author{S.~Heinz$^{1,2}$\\$^{1}$ Kavli Institute
for Astrophysics and Space research, MIT, 77 Mass. Ave., Cambridge, MA
02139\\$^{2}$ Chandra Fellow}

\begin{abstract}
We model the observed size and brightness of the VLBA radio core of
the jet in Cygnus X-1 to derive an expression for the jet power as a
function of basic jet parameters.  We apply this expression to recent
constraints on the jet power from observations of a large scale
shocked shell around the source by Gallo et al. 2005, which leads us
to a set of alternative conclusions: either (a) the jet contains large
amounts of protons: ($\geq 2000$ protons per radio emitting electron),
(b) it has a very low volume filling factor of $f \lesssim 3\times
10^{-5}$, (c) the steady, radio emitting VLBA jet is not the source of
the kinetic energy powering the ISM shell, or (d) its asymptotic
behavior differs fundamentally from a broad set of plausible analytic
jet models.
\end{abstract}

\newcommand{\sax}{SAX J1808.4-3658 }
\newcommand{\grs}{GRS 1915+105 }
\newcommand{\gro}{GRO J1655-40 }
\newcommand{\gx}{GX 339-4 }
\newcommand{\cyg}{Cygnus X-1 }

\section{Introduction}
\label{eq:introduction}
The field of X-ray binary (XRB) study has seen a remarkable
transformation over the past 5 years: Relativistic jets have moved
from being considered exotic and rare abnormalities to being
recognized as integral and maybe vital components in the transfer of
energy and angular momentum in accreting stellar mass compact objects.
While it has been known for decades that the jets produced by
supermassive black holes carry enormous amounts of energy and
transform their environment, it has been difficult to make firm
estimates of the kinetic power of XRB jets.  

Furthermore, the study of XRB jets is plagued by the same dilemma
facing extragalactic jet research for decades: a lack of solid
information about jet composition (i.e., do jets mostly contain
protons or positrons as positive charger carriers, and is the inertia
dominated by non-thermal relativistic particles or cold, thermal
plasma), jet speed (how relativistic are the jets in their very
cores), and the jet power.  These questions have been discussed for
decades \citep[e.g.][]{reynolds:96,wardle:98,ghisellini:01}, yet conclusive
observational evidence to answer them is sparse at best, so the debate
continues.  In this situation, tight constraints from individual
objects are our best bet at making progress on any of these questions.

The interest in XRB jets has arisen from the discovery of the nearly
universal presence of compact, flat spectrum radio emission from
accreting black hole X-ray binaries \citep{fender:01c} in the so-called
low/hard state, where most of the X-ray emission is in the form of a
hard powerlaw \citep[see][for a detailed review of black hole X-ray
states]{mcclintock:03}.  In two cases, this radio emission has been
resolved into clearly collimated jets \citep{dhawan:00b,stirling:01}.
The spectral and morphological similarity to the cores of AGN jets
further strengthens the interpretation of this radio emission as
evidence for jets.

However, these jets are not resolved transverse to their axes and in
the absence of other information, it has been difficult to constrain
the kinetic power of these jet - opening angles have been assumed, but
the results are then always subject to an unknown, arbitrary parameter.
Furthermore, the information encoded in the spatial distribution of the
observed radio emission of the VLBI observations of the two jets which
have been resolved (Cygnus X-1 and GRS 1915+105) have not been
harvested to a degree that allows tight constraints to be derived.
The aim of this paper is to derive just such constraints.  Cygnus X-1
is far better suited for such a study because its distance is know to
significantly higher accuracy than that of GRS 1915+105.

The status of firm observational data on XRB jet power was also
recently improved by the discovery of a shell of thermal emission
around Cygnus X-1 \citep{gallo:05}.  This shell has been interpreted as
the shocked ISM around a low surface brightness, jet-driven radio
lobe.  Using analysis borrowed from radio galaxy dynamics, the authors
limit the average kinetic power $\langle W \rangle$ from the source to
fall between $3\times 10^{36}\,{\rm ergs\,s^{-1}} < \langle W \rangle
< 3\times 10^{37}\,{\rm ergs\,s^{-1}}$.  This discovery, together with
the VLBA observations of resolved jet emission, makes Cygnus X-1 an ideal
candidate to tighten the constraints on fundamental jet parameters.

In order to put these limits into the context of the compact VLBA jet
observed by \cite{stirling:01}, we apply a representative emission
model to the VLBA data (\S\ref{sec:radio}).  This allows us to derive
a parameterized expression for the kinetic power which depends only on
a few unknowns which we hope to constrain, namely the particle content
of the jet, the filling factor of emitting material, and the
equipartition fraction of the magnetic field.  In
\S\ref{sec:discussion} we compare the outcome to the results of
\citep{gallo:05}, which is orders of magnitude larger than the
estimate we derive for our fiducial set of parameters, and discuss the
quantitative constraints that can be derived on the set of interesting
parameters from this comparison.  Section \ref{sec:summary} summarizes
our findings.

\section{The emission from the jet in Cygnus X-1}
\label{sec:radio}
The radio spectrum of the jet in Cygnus X-1 is flat to slightly
inverted ($\alpha \equiv d\log{L_{\nu}}/d\log{\nu}\gtrsim 0$), with a
flux level varying between a few and a few tens of mJy
\citep{brocksopp:99}.  \citet[][epochs A and C]{stirling:01} resolved
about 50\% of the emission to be extended, measuring roughly $\mu \sim
10-15$ mas.  The canonical model for the flat radio emission from jet
cores goes back to \citet[][BK hereafter]{blandford:79}.  This model
has been used successfully and extensively in the context of both AGN
jet cores and X-ray binary jets \citep[e.g.][]{hjellming:88,falcke:96}
and we will employ it here as well.  The underlying assumption of the
model is that the jet is described well by a freely expanding flow
with roughly uniform velocity, implying a constant half-opening angle
$\phi$.  Such a description should be a good first order approximation
at least in a limited region around the location where most of the
radio emission originates.  In such a jet, the magnetic field evolves
quickly into an almost purely toroidal topology and follows $B \propto
z^{-1}$, where $z$ is the distance along the jet, measured from the
jet origin at the compact object.

Rather than assuming adiabatic evolution of the particle distribution,
the BK model imposes fractional equipartition between particles and
magnetic fields such that the particle pressure is $p_{\rm part} =
\frac{p_{\rm mag}}{\xi_{\rm B}} \propto z^{-2}$ is a fixed fraction of
the magnetic pressure $p_{\rm mag}$, where $\xi_{\rm B}$ is the
equipartition fraction between particles and the magnetic
field\footnote{$\xi_{\rm B}$ is a free, unconstrained parameter.  The
rough proportionality between $p_{\rm part}$ and $p_{\rm mag}$ near
the radio emission region can be relaxed slightly.  However, a
strongly non-linear relation between the two would lead to steep
synchrotron spectra and is thus ruled out observationally.}.  One may
allow for the presence of protons with a contribution $p_{\rm prot} =
\xi_{\rm p} p_{\rm part}/(1 + \xi_{\rm p})$ to the particle
pressure.  The electrons (and possibly positrons, subsumed in the
non-proton part of the particle pressure) are assumed to follow a
powerlaw distribution with index $s$ such that the number density
follows $n(\gamma)=C_{\rm e}\gamma^{-s}$.  For simplicity and lack of
better measurements in the case of Cygnus X-1 we will use $s=2$, which is
the fiducial value often used in AGN jets.  We will also assume that,
due to Doppler boosting, the emission is entirely dominated by the
approaching jet, which is supported by the lack of emission detected
from the counterjet in Cygnus X-1.  Finally, we assume that the
synchrotron emitting and absorbing plasma has a volume filling
fraction of $f\leq 1$.

The synchrotron emissivity, measured in the observer's frame, is given
by
\begin{eqnarray}
j_{\nu} & = & \frac{2.4\times 10^{-17} \,{\rm ergs}}{\rm
cm^{3}\,Hz\,s} \frac{2 p^{7/4} \xi_{\rm B}^{3/4}\delta^2 f}{1 +
\xi_{\rm p}} \left[\frac{2}{1 + \xi_{\rm B}}\right]^{\frac{7}{4}}
\left[\frac{8.4\,{\rm GHz}}{\nu}\right]^{\frac{1}{2}}\nonumber \\
& \equiv & C_{0} \,
p^{\frac{7}{4}}\delta^2
\end{eqnarray}
and the self-absorption coefficient, again measured in the observer's
frame, is
\begin{eqnarray}
\alpha_{\nu} & = & \frac{2.3\times 10^{-12}}{\rm cm} \frac{2
  p^2\xi_{\rm B}}{1+\xi_{\rm p}} \left[\frac{2}{1 + \xi_{\rm
  B}}\right]^2 \left[\frac{8.4\,{\rm GHz}}{\nu}\right]^{3} \delta^2 f
  \nonumber \\ &\equiv & C_1 p^2\delta^2
\end{eqnarray}
with the obvious definitions of $C_0$ and $C_1$ \citep{rybicki:79}.
$p$ is the total (magnetic plus particle) pressure, measured in the
jet frame.  $\delta \equiv [\Gamma\left(1 -
\beta\cos{(\theta)}\right)]$ is the Doppler factor .  The jet velocity
$v \equiv \beta c \equiv c\sqrt{1-1/\Gamma^2}$ likely falls into the
range of $0.5 < \beta < 0.7$ and the viewing angle falls into the
range $25^{\circ} < \theta < 50^{\circ}$
\citep{stirling:01,brocksopp:02,gleissner:04}.  We will use $\beta
\sim 0.6$ and $\theta \sim 35^{\circ}$ as fiducial values, which gives
$\delta \equiv 2.5 \delta_{1.6}$.

For radio frequencies, the innermost region of the jet is optically
thick, not contributing significantly to the total luminosity (see
Fig.~1), which allows us to extend the lower integration limit for the
luminosity along the jet axis (denoted by the $z$-coordinate) to $z
\rightarrow 0$.  The luminosity emitted in a frame comoving with the
jet plasma is then
\begin{equation}
L_{\nu} = \int_{0}^{\infty}dz \sin{(\theta)} \int_{-\phi z}^{\phi z}dx
\frac{j_{\nu}}{\alpha_{\nu}}\left(1 - \exp{(-\tau_{\nu})}\right)
\label{eq:lumin}
\end{equation}
for the optical depth $\tau_{\nu}(z,x) = \frac{2
\alpha_{\nu}}{\sin{(\theta)}}\sqrt{\left(\phi z\right)^2 - x^2}$.  We
then define the photospheric radius $z_0$ as the location where the
optical depth through the spine of the jet is one, i.e.,
$\tau_{\nu}(z_0,x=0)=1$ and express all distances in units of $z_0$
and all other quantities relative to their value at $z_0$.  The
pressure can then be written as $p=p_0 \left({z}/{z_0}\right)^{-2}$
with
\begin{equation}
p_0=\sqrt{{\sin{(\theta)}}/\left({2 C_1 \delta^2 \phi z_0}\right)}
\label{eq:pressure}
\end{equation}
The jet luminosity from eq.~(\ref{eq:lumin}) is then
\begin{eqnarray}
L_{\nu} & = &
5.1\,{z_0}^{\frac{17}{8}}{\left[{\sin{(\theta)}}\right]}^{\frac{7}{8}}
{\phi}^{\frac{9}{8}} C_0 C_1^{-\frac{7}{8}}\delta^{\frac{1}{4}}
\label{eq:lumin2}
\end{eqnarray}
As was shown by BK, the emitted spectrum is flat, $\alpha=0$.

The VLBA flux of the Cygnus X-1 jet is $F_{\nu}\sim 12\,{\rm mJy}$
\citep{stirling:01}.  For a distance of $D \equiv 2\,{\rm kpc}\,
D_{2}$, the {\em observed} luminosity is
\begin{equation}
L_{\rm \cyg} = 5.7\times 10^{19}\,{\rm
ergs\,s^{-1}\,Hz^{-1}}\,{D_{2}}^{2}
\label{eq:lumin3}
\end{equation}

Epochs A and C are very similar in their characteristics: About 50\%
of the flux is resolved along the jet and both have essentially equal
total fluxes of $12$ and $12.9$ mJy and equal angular resolution along
the jet.  Epoch B has a lower resolved-to-unresolved flux ratio

\begin{figure}
\resizebox{\columnwidth}{!}{\includegraphics{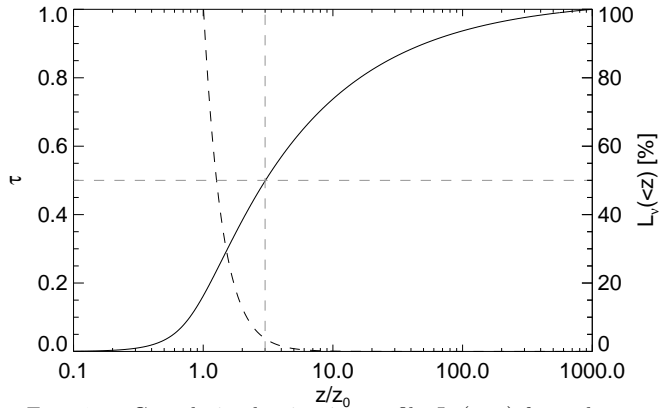}}
\caption{Cumulative luminosity profile $L_{\nu}(<z)$ from the core to
  $z/z_0$ (solid curve) and the optical depth to
  synchrotron-self-absorption across the jet at location $z/z0$
  (dashed curve). Also shown is the location of the 50\% radius,
  $z_{50\%}$ (dashed grey lines) at about $z_{50\%} \sim
  3z_0$.\label{fig:fluxprofile}}
\end{figure}

In epochs A and C of \cite{stirling:01}, 50\% of the jet flux is
resolved along the jet axis with an angular resolution of $\mu_{50\%}
\sim 3{\rm mas}$, we can go back to eq.~(\ref{eq:lumin}) and solve for
the value of $z_{50\%}$ for which 50\% of the jet emission comes from
regions with $z > z_{50\%}$.  Numerical evaluation (see
Fig.~\ref{fig:fluxprofile}) shows that this is the case for $z_{50\%}
\sim 3 z_{0}$.  Given $\mu_{50\%} \sim 3{\rm mas}$, we can thus
determine the photospheric distance $z_0$:
\begin{equation}
z_0\sim 5.2\times 10^{13}\,{\rm
cm}\,\left(\frac{\sin{35^{\circ}}}{\sin{\theta}}\right)\left(\frac{\mu_{\rm
50\%}}{3{\rm mas}}\right) D_2
\label{eq:corelength}
\end{equation}
Epoch B (which has slightly lower resolution along the jet) shows a
more compact jet, with only about 30\% of the flux resolved, which
would imply a somewhat lower value of $z_0$ (by about a factor of 2).
This might be due to changes in viewing angle due to jet precession,
or possible interaction with the powerful wind from the companion (see
\S\ref{sec:discussion}).  We will thus carry the dependence on
$\mu_{50\%}$ through for transparency.  Strictly speaking,
eq.~(\ref{eq:corelength}) is a lower limit on $z_{0}$, since the VLBA
observations might have resolved out and thus missed some of the
larger scale, low surface brightness emission.  However, since the
total flux in the VLBA image is close to the total flux seen by other
instruments, it is unlikely that $z_0$ is off by more than a few tens
of percent.

It is noteworthy that, from eq.~\ref{eq:lumin2}, the flux density is
\begin{equation}
F_{\nu} = L_{\nu}/(4\pi D^2) \propto
\left(\mu_{50\%}\right)^{\frac{17}{8}} D^{\frac{1}{8}}
\left(\sin\theta\right)^{-\frac{10}{8}}\delta^{\frac{1}{4}}
\end{equation}
i.e., for a given angular size $\mu_{50\%}$, the flux from a compact
jet is {\em almost} independent of the distance $D$ to the source!
Because the angular distance and the luminosity distance scale
differently with redshift, it should be possible in principle to use
jet cores as yard sticks.  The main obstacle for such a use is the
fact that relativistic beaming will introduce significant uncertainty
and bias, such that a statistical analysis of this effect will almost
certainly be flawed.  However, given a distance estimate, a
measurement of $F_{\nu}$ and $\mu_{50\%}$ will yield a constraint on
$\left(\sin\theta\right)^{-\frac{10}{8}}\delta^{\frac{3}{8}}$.
Together with independent constraints on $\theta$ (e.g., from the
jet-counterjet surface brightness ratio), this provides a constraint
on the jet velocity, with very little error associated with distance
uncertainties, which are dominant in velocity measurements based on
superluminal proper motion \citep{fender:03b}.

By equating eqs.~(\ref{eq:lumin2}) and (\ref{eq:lumin3}) and
substituting from eqs.~(\ref{eq:pressure}) and (\ref{eq:corelength})
we can calculate the jet opening angle:
\begin{eqnarray}
  \phi & \sim & 0.06^{\circ} \left[\frac{\sin{\theta}}
  {\sin{35^{\circ}}}\right]^{\frac{10}{9}} \left[\frac{1 +
  \xi_{\rm p}}{\delta_{1.6}^{2}2fD_2}\right]^{\frac{1}{9}} \nonumber
  \\
  & & \ \ \ \ \ \ \ \ \ \ \ \times \left[\frac{3{\rm
  mas}}{\mu_{50\%}}\right]^{\frac{17}{9}}\left[\frac{F_{\rm
  8.4GHz}}{12{\rm mJy}}\right]^{\frac{8}{9}}
  \label{eq:opening}
\end{eqnarray}
The value we find for $\phi$ is very small compared to jet opening
angles observed in AGN jets (e.g., in the M87 jet, the half-opening
angle of the large scale jet is of order $\phi \sim 1.5^{\circ}$), but
entirely consistent with the upper limit of $\phi < 2^{\circ}$ set by
the VLBA observations \cite{stirling:01}.  If the jet is in free
expansion, this small opening angle would imply a very low internal
sound speed, i.e., a pool of cold particles in addition to the
synchrotron emitting electrons (see \S\ref{sec:discussion}).  Note
that the value of $\phi$ is rather sensitive to $\mu_{50\%}$, which
introduces significant uncertainty given the variance of this value
between epochs A through C of \cite{stirling:01}, up to a factor of 5
(the largest value of $\phi\sim 0.3^{\circ}$ is found for epoch B).

Inserting this into eq.~(\ref{eq:pressure}) we get
\begin{eqnarray}
  p_0 & \sim & \frac{1\,{\rm erg}}{\rm cm^{3}} \frac{1+\xi_{\rm
  B}}{2\delta_{1.6}^{\frac{8}{9}}\xi_{\rm B}^{\frac{5}{9}}}
  \left[\frac{\sin{\theta}}{\sin{35^{\circ}}} \frac{1+\xi_{\rm
  p}}{2fD_{2}} \frac{\mu_{50\%}}{3 \,{\rm mas}} \frac{12\,{\rm
  mJy}}{F_{8.4{\rm GHz}}}\right]^{\frac{4}{9}}
\end{eqnarray}

Given the opening angle and pressure of the jet, we can now estimate
the kinetic jet power for the leptonic and electromagnetic jet content
(i.e., not counting the rest mass kinetic energy in possible thermal
particles the jet might be carrying, but including both jet and
counterjet):
\begin{eqnarray}
  W_{\rm min} & \sim & 2 \times 4p\Gamma^2 \beta c \pi (\phi\,z)^2
  \nonumber \\ &\sim & 1.9 \times 10^{33}\,{\rm ergs\,s^{-1}}
  \left(\frac{2}{1+\xi_{\rm
  B}}\right)^{\frac{5}{9}}\Gamma_{1.25}^2\beta_{0.6} \nonumber \\ & &
  \left[\frac{\sin{(\theta)}}{\sin{(35^{\circ})}}\left(\frac{12\,{\rm
  mJy}}{F_{8.4{\rm GHz}}}\frac{3\,{\rm mas}}{\mu_{50\%}}\right)^{2}
  \frac{\xi_{\rm B}\left(1 + \xi_{\rm
  P}\right)}{2fD_{2}^2\delta_{1.6}^2}\right]^{\frac{2}{3}}
  \label{eq:power1}
\end{eqnarray}
This estimate is significantly smaller than the $W \sim 10^{35}\,{\rm
ergs\,s^{-1}}$ estimated by \cite{spencer:01}\footnote{semi-analytic
models by \cite{markoff:01} also give significantly larger values for
the power.} and typical values assumed in the literature for compact
XRB jets \citep{fender:03}.  This is mainly due to the smaller opening
angle required by the observed value of $\mu_{50\%}$.  Unlike the
value for $\phi$ from eq.~(\ref{eq:opening}, the $W$ estimate in
eq.~(\ref{eq:power1}) is less sensitive to the value of $\mu_{50\%}$.
The $\mu_{50\%}^{4/3}$ dependence still introduces an uncertainty of
up to a factor of 3, given the variance in $\mu_{50\%}$ and
$F_{8.4GHz}$ seen between the three different epochs of
\cite{stirling:01}, with epoch B leading to a power estimate of
$W_{\rm min} \sim 6\times 10^{33}\,{\rm ergs\,s^{-1}}$.

\section{Discussion} 
\label{sec:discussion}
At face value, this result implies that the kinetic jet power carried
by fields and radio emitting electrons is smaller than the radiative
power in X-rays by about 4 orders of magnitude.  If other X-ray binary
jets in the low--hard state were similar in nature to the jet in \cyg,
this would imply that jets do {\em not} carry away most of the
accretion power in low--hard state and quiescent sources, as recently
suggested by \cite{fender:03}.  This would require an inefficient
accretion scenario like an ADAF \citep{narayan:94}, CDAF
\citep{quataert:00}, or ADIOS \citep{blandford:99}.  It would also
imply that recent, independent estimates of the jet power from steady
low/hard state jets \citep{fender:05b,heinz:05c} are off by four
orders of magnitude.

Such a low power outflow in Cygnus X-1 is in {\em direct
contradiction} to the energy requirement for the recently found
thermal shell around the putative radio lobe of Cygnus X-1
\citep{gallo:05}, which requires an average jet power of
\begin{equation}
  3\times 10^{36}\,{\rm ergs\,s^{-1}} < W_{\rm shell} < 3\times
  10^{37}\,{\rm ergs\,s^{-1}}
  \label{eq:powershell}
\end{equation}

Furthermore, several low/hard state XRBs show IR excesses that have
been convincingly interpreted as jet emission (XTE J1118+480,
\citealt{fender:01b}, GX 339-4, \citealt{homan:05,nowak:05}, GRS
1915+105, \citealt{ogley:00}) with total radiative energy output from
the jet in excess of 5\% of the total radiative output, amounting to
$L_{\rm jet} > 2\times 10^{35}\,{\rm ergs\,s^{-1}}$ in the case XTE
J1118+480 \citep{fender:01b}.  The jet power must exceed the radiative
power by a significant margin since otherwise the jet would radiate
away more energy than it is carrying and thus violate energy
conservation\footnote{Even the X-ray emission in the low/hard sate
systems might partially originate in the jet
\citep{markoff:01,markoff:04}, in which case the power requirement
would increased by another 2 orders of magnitude, though synchrotron
emission is probably not the dominant source of the X-rays
\citep{heinz:04a}.}.  While the bright companion of Cygnus X-1 makes
direct observations of an IR/optical contribution by the jet
difficult, \cite{fender:01} argues that a flat jet spectrum extending
to the V-band would still require a kinetic power in excess of $\sim
10^{35}\,{\rm ergs\,s^{-1}}$.

It is thus rather unlikely that the power of the Cygnus X-1 jet is
truly as low as implied by eq.~(\ref{eq:power1}) for the fiducial
choice of parameters.  We will discuss four alternative (not
necessarily mutually exclusive) conclusions that can be reached from
the mismatch between eqs.~(\ref{eq:power1}) and (\ref{eq:powershell}):

\subsection{Jet composition}
\label{sec:coldmatter}
It is clear from eq.~(\ref{eq:power1}) that $W_{\rm min}$ depends on
$\xi_{\rm B}$ and $\xi_{\rm p}$, thus, jet composition and field
strength both factor into the total power.  

It is straight forward to show that the a significant deviation from
equipartition between particles and magnetic field will {\em not}
change the estimate of eq.~(\ref{eq:power1}) anywhere close to what is
needed to bring it in line with the limits from \cite{gallo:05}.  The
asymptotic dependence on $\xi_{\rm B}$ is as follows: for small
$\xi_{\rm B}$, $W_{\rm min}$ actually {\em decreases} rather than
increasing, thus exacerbating the power discrepancy.  For large
$\xi_{\rm B}$ on the other hand, $W_{\rm min} \propto \xi_{\rm
B}^{1/9}$, requiring $\xi_{\rm B} \sim 10^{36}$ to bring $W_{\rm min}$
in line with the results by \cite{gallo:05}.  Such large values are
clearly unreasonable and unphysical.  In other words: we cannot
constrain the magnetic field strength, but significant deviations from
equipartition will also {\em not} bring eq.~(\ref{eq:power1}) in line
with eq.~(\ref{eq:powershell}).

On the other hand, if the jet contains a large amount of cold, thermal
plasma, the kinetic power can be significantly enhanced.  Before going
into more detail, it is important to note that simple charge balance
of cold protons with the observed synchrotron emitting electrons will
not enhance the power by more than two orders of magnitude because the
above estimate is based on minimum energy arguments.  For $\xi_{\rm B}
\sim 1$, the Lorentz factor of the electrons contributing most of the
8.4 GHz flux is about $\gamma_{\rm e,8.4GHz} \approx
70\left(p_{0}/1\,{\rm erg\,cm^{-3}}\right)$, thus adding one cold
proton per emitting electron would only raise the kinetic power by
about an order of magnitude, insufficient for bringing it in line with
eq.~(\ref{eq:powershell}).

The kinetic jet power including the inertial term is
\begin{equation}
  W = W_{\rm min} + 2 \rho c^3 \Gamma\beta\left(\Gamma -
  1\right)\pi\left(\phi z\right)^2
\end{equation}
where $w=\gamma/(\gamma - 1)p + \rho c^2$ is the enthalpy.  The amount
of cold gas necessary to bring the power estimate in line with the
large scale constraints from eq.~(\ref{eq:powershell}, parameterized as
$W_{\rm tot}=3\times 10^{36}\,{\rm ergs/s}W_{36.5}$, by
\cite{gallo:05} is then
\begin{equation}
  \rho_0 \gtrsim \frac{2.4\times 10^{-17}\,{\rm g}}{\rm
  cm^{3}}\left[\frac{0.06^{\circ}}{\phi}\right]^{2}
  W_{36.5}\frac{3/16}{\Gamma\beta(\Gamma-1)}
  \label{eq:rhomin}
\end{equation}
or about 2000 protons per radio emitting electron (we have abbreviated
the dependence on the underlying parameters $\xi_{\rm B}$, $\xi_{\rm
p}$, and $f$ in the opening angle $\phi$ - see eq.~(\ref{eq:opening})
for the detailed parametric dependence).

This is a firm lower limit.  We should note again that the question
whether jets contain large amounts of protons has been discussed in
the literature many times over but remains essentially unanswered to
this day, especially in the newer and less well studied class of XRB
jets.  Any new constraint on an individual object, like the one
derived here, will thus be valuable addition to the debate.  Note also
that energetic requirements for the presence of cold, thermal protons
(derived here) are different from the requirement of a thermal parent
population of leptons at the base of the jet, as required in models
that attempt to match not only the radio but also IR through X-ray
observations of X-ray binaries, such as those by
\citep{markoff:01,markoff:04,yuan:05}.  These particles are required
on spectral grounds, not for energetic reasons and no quantitative
requirement for the presence of cold protons can be made from them.

Arguments for the presence of cold protons {\em have} been made on
dynamical grounds in order to explain the opening angles of jets in
the case of freely expanding, ballistic jet models
\citep{falcke:96,markoff:01}.  However, the opening angle and the jet
power in those models are free parameters and thus constraints on the
presence of protons in the jets are qualitative {\em only}. We will
employ a similar argument in the following paragraphs, based on the
new constraint we derived for the jet opening angle in
eq.~(\ref{eq:opening}).

Above and beyond the lower limit from eq.~(\ref{eq:rhomin}), the total
amount of cold plasma that might be traveling down the jet cannot be
constrained in the absence of direct radiative signatures, and thus
the power enhancement that could be provided by cold particles is
formally unconstrained.  However, the small opening angle implied by
eq.~(\ref{eq:opening}) could be interpreted as evidence for the
presence of large amounts of cold matter if we assume that the jet
opening angle is approximately equal to the Mach cone of the
jet\footnote{Note that in a BK type jet the ``temperature'' ($p/\rho$)
of the relativistic plasma is constant along the jet, thus defining a
temperature based on the jet opening angle is well posed and
meaningful.  However, even if the thermal plasma were not of constant
temperature along the jet because of radiative cooling, this argument
would still hold, as the opening angle would define the temperature at
the location where the jet becomes ballistic (i.e., inertia
dominated).  Further out, the thermal temperature might be lower than
at that point, but the opening angle would remain at its ballistic
value.}, i.e.,
\begin{equation}
  \phi \sim c_{\rm s}/({\beta c \Gamma}) \sim \sqrt{{5
  p}/{3\rho}}/({\beta c \Gamma})
  \label{eq:phi}
\end{equation}
Thus, if cold, thermal particles, such as protons (with the
appropriate number density of cold electrons for charge balance), are
present in the jet and are the reason for the small opening angle, we
can use the estimate of $\phi$ and eq.~(\ref{eq:phi}) to obtain an
estimate of the rest mass density $\rho$ inside the jet.  This is, in
effect, an upper limit because the jet might be partially collimated
by magnetic fields and thus not in free, ballistic expansion.
\begin{eqnarray}
  \rho_0 & \lesssim & \frac{3.2\times 10^{-15}\,{\rm g}}{\rm
  cm^{3}}\frac{p_0}{1\,{\rm
  ergs\,cm^{-3}}}\left(\frac{0.06^{\circ}}{\phi}
  \frac{0.75}{\Gamma\beta}\right)^{2}
\end{eqnarray}

The jet power is then
\begin{eqnarray}
  W & \lesssim & 3 \times 10^{38}\,{\rm
  ergs\,s^{-1}}\frac{\rho_0}{4.5\times10^{-16} {\rm g
  cm^{-3}}}\frac{\Gamma\beta(\Gamma - 1)}{3/16}\ \
  \label{eq:power2}
\end{eqnarray}
In this case, the high particle densities of $n\sim 2 \times
10^{9}\,{\rm cm^{-3}}$ might imply a detectable thermal X-ray flux
from these particles (as is the case in the jet of SS433).  The
temperature of this thermal gas would be $T=2.3 \times 10^6\,{\rm
K}(\phi/0.06^{\circ})^2(\beta\Gamma/0.75)$.  The total thermal 0.5-10
keV luminosity from the jet would be $L_{\rm brems} \approx 8\times
10^{37}\,{\rm ergs\,s^{-1}}$, comparable to the bolometric luminosity
believed to be coming from the accretion flow.  The associated soft
X-ray/UV/optical line emission, which would be very similar to the
spectrum of SS433, has not been observed.

Such a dense plasma would cool rapidly\footnote{It should be noted
that a fixed opening angle in a cooling jet is {\em not} a
contradiction: the opening angle of a ballistic jet is set at the
point where the jet goes out of transverse causal contact, i.e., where
the sideways expansion becomes faster than the effective internal
sound speed of the jet plasma (e.g., the fast magnetosonic speed in a
magnetized jet).  Beyond that point, the jet will expand at an
essentially fixed rate, possibly cooling adiabatically and/or
radiatively.  In other words, the temperature does not need to be
proportional to $\phi^2$ everywhere along the jet.  Thus, contrary to
a first glance educated guess, no mechanism is necessary to reheat the
plasma to maintain a fixed opening angle.} if it were present from the
very base of the jet (the cooling time there would be about $10^{-9}$
seconds, much shorter than the dynamical time).  In order to avoid
catastrophic radiative cooling, particles would have to injected at a
large distance from the black hole, $z_{\rm cool}\gtrsim 10^{13}\,{\rm
cm}$ (where the radiative cooling time is longer than the dynamical
time).  Coincidentally, this is comparable to the orbital separation
of $\sim 10^{12}\,{\rm cm}$ of the system.

It would be natural, then, to interpret the presence of such a
putative thermal component as mass loading from the powerful wind
blown by the companion.  Since this thermal plasma is supposed to set
the opening angle of the jet, collimation and acceleration would still
have to be happening at the location where the mass loading takes
place, which is very far from the black hole.  This would provide
interesting asymptotic constraints for models of jet dynamics.

\subsection{Low filling factor}
\label{sec:filling}
Another way to increase the power estimate from eq.~(\ref{eq:power1})
would be a filling factor $f$ much smaller than one. While the opening
angle $\phi$ depends only weakly on $f$, the jet power is somewhat
more sensitive.  A very low filling factor would translate to a
physical picture where small regions dissipate a large amount of
energy into relativistic particles, such as would occur in magnetic
reconnection events in a turbulent flow.  

In this context, it is interesting to note that high resolution radio
images of AGN jets (which have are much better suited for high
resolution imaging due to their larger ratio of object size to
distance) show very non-uniform and sometimes unresolved, filamentary
emission at least in the regions where the jet is optically thin
\citep[e.g.][]{biretta:95}.  It might well be that the emission from the
smaller, optically thin region is similarly filamentary.

In order to bring the estimate of eq.~(\ref{eq:power1}) back in line
with the macroscopic limits \citep{gallo:05}, the filling factor would
have to be $f \lesssim 3\times 10^{-5}$, which is very small indeed
and difficult to achieve with internal shock scenarios \citep{yuan:05}
or multi-zone continuous emission models \citep{markoff:01}.

\subsection{A non-radiative source for the kinetic power}
It is also possible that the flat spectrum, steady, low/hard state jet
that is observed directly in the VLBA radio images is {\em not} the
source of the power that drives the ISM shell.

For example, it is possible that the radio emission originates in a
spine along the jet axis, which is embedded in a much wider sheath of
low emissivity plasma with opening angle $\phi_{\rm sheeth}$. In this
case, the true kinetic power could be much larger (by a factor of
$\left(\phi_{\rm sheeth}/\phi\right)^2$.  To bring the power estimate
in line with the value from \cite{gallo:05}, we would require
$\phi_{\rm sheeth} \geq 2^{\circ}$.  Note that this is different from
a low filling factor scenario since the optical depth through the
spine is larger than that through a low filling factor jet of equal
emission measure.  Why most of the flow would be non-radiative in this
case is similarly difficult to understand as the case of a low filling
factor, however.  It would amount to postulating an unrelated, much
larger angle flow that would supply all the power but no radiation
and, in essence, would mean that the actually observed jet has no
physical relation to the large scale bubble observed around Cygnus
X-1.  Whether a radiative signature of such a larger angle outflow
exists and how it would tie in with ADIOS and disk wind models is not
clear.  

Since such a putative outflow has so far not been observed, there are
no constraints on the nature of the flow other than the kinetic power
one would have to postulate based on the large scale requirements.  It
would thus be impossible to derive mass flow rates and outflow
velocities.  However, given the observed power and using the Eddington
accretion rate of $\dot{M} = 1.5\times 10^{19}\,{\rm
g/s}(M/10M_{\odot})$ for a 10 solar mass black hole as an upper limit
on the mass flow rate, we can conservatively put a lower limit of
$v_{\rm sheeth} > 6000\,{\rm km\,s^{-1}} W_{36.5}^{1/2}$ on the
outflow velocity.

It is also possible that most of the kinetic power in the Cygnus X-1 jets
is supplied by the type of optically thin radio outbursts observed in
many Galactic XRB sources like, for example, GRS 1915+105
\citep{mirabel:94,fender:99} rather than the steady compact jets
discussed here.  This would have interesting consequences for the
estimates of duty cycles and radiative efficiencies for such events.
Since they are unrelated to the VLBA observations discussed here, we
will not discuss this possibility in more detail.

It should be noted that the flat spectrum IR emission from low/hard
state XRBs that has been interpreted successfully as jet emission
still places power constraints on the jet that, while smaller than
eq.~(\ref{eq:powershell}), still far exceed that of
eq.~(\ref{eq:power1}).  IR observations of Cyg X-1 are difficult
because of the high Galactic background and the bright companion, thus
we cannot apply these limits directly.  However, if the Cyg X-1 jet
exhibits IR fluxes similar to those from other low/hard state jet
sources, an unrelated source of kinetic power that might be
responsible for the large scale ISM shell could not explain the tight
IR/radio correlation, and thus the constraints derived in
\S\ref{sec:coldmatter} and \ref{sec:filling} would hold.  Ongoing
Spitzer IR monitoring will help to answer this question.

\subsection{Grossly different kinematics}
Finally, the treatment used in this paper might not be applicable
because some of the basic assumptions we employed might be false.
This would imply that the most commonly used description for jet cores
that captures the critical observable features, the
``Blandford/Koenigl'' model, would grossly misrepresent the kinematics
and emissivity characteristics of jet cores (by many orders of
magnitude).  This in itself would be an important constraint on jet
modeling.  We shall briefly consider whether a :

The most obvious deviations from a BK model would be (a) a non-conical
jet, (b) a deviation from the assumed $p\propto z^{-2}$ scaling, or
(c) an electron spectral index $s$ different from $s=2$.  Barring
seriously pathological jet geometries and kinematics, the jet should
be well described (at least near the radio photosphere) by powerlaws
for the jet radius $R = R_0(z/z_0)^{\zeta}$ (with $0 \leq \zeta \leq
1$ for collimation) and the pressure $p=p_0(z/z_0)^{\chi}$.  Following
eq.~(\ref{eq:power2}), the kinetic power of the jet is $W =
\left[4p\Gamma^2\beta c + \rho\left(\Gamma^2 - \Gamma\right)\beta
c^3\right]\pi R^2$.  If the jet is {\em not} cold matter dominated
($\rho c^2 \ll p$), the energy flux is only constant along the jet if
$\chi=-2\zeta$, which we impose as a condition in the following.  If
it {\em is} cold matter dominated, once again \S\ref{sec:coldmatter}
applies.

The spectral index for such a jet would be $\alpha = (18\zeta - 8 - 2s
- 3s\zeta)/12\zeta$.  Since the observed spectrum in low/hard state
XRB jets is flat to slightly inverted (in the case of GX 339-4,
$\alpha \sim 0.1$, while at low frequencies, the jet in XTE J1118+480
is rather steep, $\alpha \lesssim 0.5$), we can impose $\alpha \geq 0$
and find $s \leq (18\zeta - 8)/(2 + 3\zeta) \leq 2$.  Fermi
acceleration, which is often invoked as the origin of the electron
powerlaw spectra, typically produces spectra with $s \geq 2$, which is
consistent with the above condition {\em only} for the case $s=2$ and
$\zeta=1$, {\em i.e.}, a BK jet.  Slightly smaller values of $s$ {\em
  might} be possible to achieve with relativistic Fermi acceleration
\citep{kirk:89} or reconnection, but the parameter space is severely
limited to a region around the classical BK model ($s=2$ and
$\zeta=1$).  Estimates presented in this paper will not be affected
too strongly by such a small deviation in parameters.

One might also argue that the assumption of uniform velocity is
incorrect, i.e., $\Gamma$ might be increasing significantly inside the
radio emission region through active hydrodynamic or MHD driving, in
which case a simple kinematic model would be inappropriate.  It is
beyond the scope of this paper to present dynamical jet models in any
detail.  However, it is clear that the radio emission region is very
far away from the black hole, roughly $10^7$ gravitational radii
according to eq.~(\ref{eq:corelength}).  Jet acceleration at such
distances would be very surprising indeed, presenting serious
asymptotic challenges to dynamical jet models.  Significant
acceleration over such large scales would also imply enormous
terminal Lorentz factors which would be at odds with the observational
velocity constraints on Cygnus X-1 \citep{gleissner:04}.

Thus, it is not trivial to construct a jet model that produces a flat
spectrum, obeys energy conservation, has reasonable terminal
velocities, and still deviates significantly from a BK-jet.

However, it is worth keeping in mind that, while the Cygnus X-1 jet is
generally well described by a BK jet, it does exhibit some features
that go beyond this simple, robust picture: The jet does show signs of
curvature in epoch A of \cite{stirling:01}, which might be due to
precession induced by the binary orbit, possibly explaining the
different appearance of epoch B compared to epochs A and C.
Furthermore, other Galactic jet sources show an even broader variety
of behaviors, such as spectral pivoting \citep{corbel:00} and short
term variability.  Such behavior clearly requires more complex models,
including the possibly confining and contaminating presence of stellar
winds (which, interestingly, should affect the Cygnus X-1 jet most
strongly, given the powerful wind from it O9.7 Iab companion).
Finally, the question of what accelerates the relativistic electrons
inside the jet (as postulated by all jet emission models), which
directly relates to the assumed quasi-equipartition between particles
and magnetic field in the BK model, remains unanswered.

It is beyond the scope of this paper to present models addressing
these issues, but it is possible that more complex dynamical models,
incorporating sophisticated particle transport calculations, will lead
to a significantly different power estimate that is more in line with
the observed values from eq.~(\ref{eq:powershell}).  As it stands, the
requirement to satisfy the observed flat spectrum and the VLBI surface
brightness distribution (in detail), while at the same time satisfying
the large scale power requirements, presents a benchmark for new jet
models to measure up to.

\section{Summary}
\label{sec:summary}
We presented a detailed analytic model of the VLBA emission from the
Cygnus X-1 radio jet to derive a parametric expression of the jet
power $W_{\rm jet}$.  By applying this expression to new, independent
observational constraints on $W_{\rm jet}$ by \cite{gallo:05}, we
derived stringent constraints on fundamental parameters of the jet,
which lead us to the following set of alternative conclusions: The jet
must either (a) contain large amounts of cold protons and/or (b) have
am extremely low filling factor, and/or (c) the power that drives the
ISM shell is not carried by the compact VLBA radio jet but some other,
unrelated type of outflow (e.g., optically thin radio flares or a
broad accretion disk wind) and/or (d) a strong asymptotic deviation
from the analytic Blandford-Koenigl model we employed (however, still
subject to the very restrictive spectral and imaging constraints
provided by the data).  Each of these alternatives presents valuable
input for jet modeling efforts.

\acknowledgements{We would like to thank Sera Markoff, Mike Nowak, and
  Rob Fender for helpful discussions.  Support for this work was
  provided by the National Aeronautics and Space Administration
  through Chandra Postdoctoral Fellowship Award Number PF3-40026
  issued by the Chandra X-ray Observatory Center, which is operated by
  the Smithsonian Astrophysical Observatory for and on behalf of the
  National Aeronautics Space Administration under contract
  NAS8-39073.}

\end{document}